\documentclass[
  reprint,        % Two-column layout (journal style)
  aps,            % American Physical Society
  pra,            % Journal style: prl, pra, prb, prc, prd, pre
  superscriptaddress,
  floatfix
]{revtex4-2}

\setcitestyle{super}

\usepackage{amsmath, amssymb}
\usepackage{graphicx}
\usepackage{booktabs}
\usepackage[dvipsnames]{xcolor}
\usepackage{calc}
\usepackage[english]{babel}
\usepackage[
  colorlinks=true,
  linkcolor=RoyalBlue,
  citecolor=BrickRed,
  urlcolor=MidnightBlue
]{hyperref}

\newcommand{\beginsupplement}{%
  \setcounter{section}{0}
  \setcounter{figure}{0}
  \setcounter{table}{0}
  \setcounter{equation}{0}
  \renewcommand{\thesection}{S\Roman{section}}
  \renewcommand{\thefigure}{S\arabic{figure}}
  \renewcommand{\thetable}{S\Roman{table}}
  \renewcommand{\theequation}{S\arabic{equation}}
}

\newcommand{\dif}{\mathrm{d}} % Defined by Klaus Jaeger

\begin{document}
% -------------------------------------------------

% -------------------------------------------------
% Title information
% -------------------------------------------------
\title{How do sub-bandgap reflectors affect the performance of photovoltaic modules?}
%\subtitle{Optional Subtitle}

\author{Klaus Jäger}
\email{klaus.jaeger@helmholtz-berlin.de}
\affiliation{Department Optics for Solar Energy, Helmholtz-Zentrum Berlin für Materialien und Energie GmbH, Kekuléstraße 5, 12489~Berlin, Germany}
\affiliation{Computational Nano Optics, Zuse Institute Berlin, Takustraße 7, 14195 Berlin, Germany}
\affiliation{Andlinger Center for Energy and the Environment, Princeton University, Princeton, New Jersey 08544, USA}

\author{Jyotirmoy Mandal} 
\affiliation{Department of Civil and Environmental Engineering, Princeton University, Princeton, New Jersey 08544, USA}

\author{Barry P. Rand}
\affiliation{Andlinger Center for Energy and the Environment, Princeton University, Princeton, New Jersey 08544, USA}
\affiliation{Department of Electrical and Computer Engineering, Princeton University, Princeton, New Jersey 08544, USA}

\author{Forrest Meggers}
\affiliation{Andlinger Center for Energy and the Environment, Princeton University, Princeton, New Jersey 08544, USA}
\affiliation{School of Architecture, Princeton University, Princeton, New Jersey 08544, USA}

\author{Christiane Becker}
\email{christiane.becker@helmholtz-berlin.de}
\affiliation{Department Optics for Solar Energy, Helmholtz-Zentrum Berlin für Materialien und Energie GmbH, Kekuléstraße 5, 12489~Berlin, Germany}
\affiliation{Faculty 1: School of Engineering – Energy and Information, Hochschule für Technik und Wirtschaft Berlin, Wilhelminenhofstraße 75A, 12459 Berlin, Germany}

%\date{\today}
\date[]{Published in \textit{Sustainable Energy \& Fuels}, 2026,
  \href{https://doi.org/10.1039/D6SE00463F}{doi:10.1039/D6SE00463F}}

\begin{abstract}
Sub-bandgap reflectors (SBR) can reduce the temperature of photovoltaic (PV) modules by reflecting the near-infrared region of the solar spectrum with photon energies smaller than the electronic bandgap of the solar cell absorber material. We consider an ideal SBR, which reflects 100\% of non-harvestable low-energy photons but does not alter the reflectivity of the PV module for usable high-energy photons, and estimate how reducing the module temperature with the SBR affects the annual and the cumulative energy yield of silicon PV modules for six locations in North America and Europe. An ideal SBR would increase the annual energy yield between 0.9\% and 1.4\% for open-rack mounted modules and between 1.5\% and 2.3\% for close-roof mounted PV modules. Whether a non‑ideal SBR provides a benefit in actual deployments strongly depends on the location and the optical properties of the coating. Beyond effects on the instantaneous power conversion efficiency and hence the annual energy yield, reducing the temperature by an SBR might also reduce the degradation and increase the overall lifetime of the PV module. By describing degradation using a simple Arrhenius approach with typical activation energies between 0.4~eV and 0.8~eV, we find that an ideal SBR increases the cumulative energy yield over 30 years between 2.2\% and 4.0\% for an open-rack mounted PV module in Princeton, New Jersey, USA.
\end{abstract}

\maketitle

\section*{Introduction}

Elevated temperatures reduce the power conversion efficiency (PCE) and the lifetime of photovoltaic (PV) modules. As a result the cumulative energy yield from PV modules over their lifetime is substantially reduced, especially in warmer climates. Typical temperature coefficients for power are between $\beta_P = -0.25$ and $-0.5~\%\,\text{K}^{-1}$ for silicon PV, meaning that the PCE of a PV module decreases by a relative  $0.25\%$ to $0.5\%$ per kelvin increase in temperature.\cite{Paudyal2021} \emph{Thermal management} approaches aim to keep PV modules and hence solar cells cooler during operation with the goal to increase energy yield and to reduce degradation. Further, PV modules can increase the temperature in urban environments.\cite{Khan2024, Hosseini2025} Reducing PV module temperatures may therefore mitigate this urban temperature rise.

Various cooling techniques have been proposed and tested.\cite{Sato2019,Dwivedi2020,KozakJagiela2023} \emph{Active cooling} approaches use \emph{forced convection} via fluids, such as air  or water, and the heat transferred to the fluid may be used in a combined photovoltaic/thermal (PV/T) system.\cite{Teo2012,MazonHernandez2013,Wu2011,Fakouriyan2019,Arcuri2014}
\emph{Passive cooling} approaches can be realized for example with \emph{convective heat transfer}\cite{Bashir2025} or \emph{phase-change materials} (PCM) that buffer heat via melting and freezing.\cite{OrtizLizcano2024} 

More recently,  \emph{radiative cooling}, a passive optical approach, has received increased attention. The atmosphere of the clear sky is quite transparent to radiation with wavelength ($\lambda$) between 8 and 13~\textmu m---the so-called atmospheric window---allowing for heat transfer with space at 2.7~K temperature. Several approaches have been investigated to maximize the thermal emissivity of PV modules in the infrared ($\lambda>4$~\textmu m).\cite{Sato2019,Zhao2018,Akerboom2022,Li2017,An2019} However, standard solar-module glass emits already more than 75\% for $\lambda>4$~\textmu m and therefore further increasing the emissivity in that wavelength range would only allow for small reductions of the module temperature.\cite{Sun2017,Li2017}

Another optical cooling approach is the use of \emph{sub-bandgap reflectors} (SBR), which reduce the absorbed heat instead of enhancing heat emission: For silicon with a bandgap of about $E_g =1.1$~eV, corresponding to $\approx1100$~nm wavelength, about 20\% of the incident solar irradiance cannot be used because the photon energy is smaller than the bandgap and therefore too small to generate electron-hole pairs. If these photons are absorbed elsewhere in the PV module, their energy will heat up the module. Hence, maximizing the sub-bandgap reflectivity of the solar module keeps the module temperature lower during operation.\cite{Sun2017, Li2017, Perrakis2020, Li2021} 
Li \emph{et al.} report a temperature reduction of 3.2--5.7~\textdegree C for an SBR, significantly larger than the 1.2--2.8 \textdegree C they report for radiative cooling.\cite{Li2021} Sun \emph{et al.} find the same trend for silicon modules, with a temperature reduction of about 4~\textdegree C for an SBR and about 2~\textdegree C for radiative cooling.\cite{Sun2017} To maximise the sub-bandgap reflection, spectrally selective mirrors were investigated.\cite{Li2017,Slauch2021,Cote2021}

To assess the real-world benefit of thermal management, the annual energy yield (AEY) must be considered. Slauch and coworkers calculated that a spectrally selective mirror would lead to 0.5--1.0~\textdegree C average temperature reduction corresponding to 0.2--0.4\% AEY increase. Because their mirrors also increase the current density, they see up to 4\% overall EY increase.\cite{Slauch2019}  Ortiz-Lizcano \emph{et al.}\ designed a multilayer stack that lead to temperature reductions of 2.20~\textdegree C and 2.45~\textdegree C for Delft, the Netherlands, and Singapore, respectively.\cite{Lizcano2024} While the initial annual energy yield was decreased by almost 10\%, they expect a benefit over the lifetime of the PV modules because of decelerated degradation.

\begin{figure}
    \centering
    \includegraphics{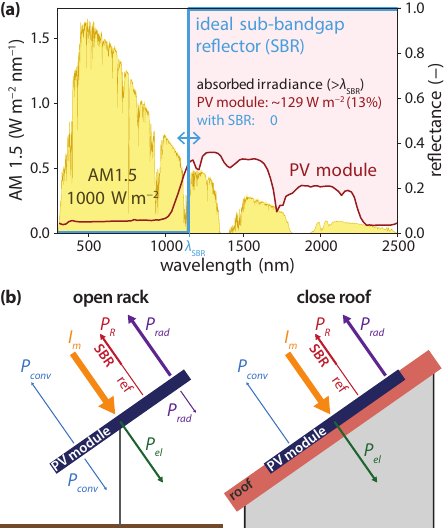}
	\caption{\label{fig:ideal} (a) AM1.5 reference solar irradiance spectrum\cite{astm:2008} (left $y$-axis), and reflectance spectra of a PERC solar module, digitized from Fig.~5b of Ref.~\cite{Subedi2020} (adapted with permission from Elsevier), and an ideal sub-bandgap reflector (right $y$-axis). The irradiance absorbed by the PV module is shown for $\lambda>\lambda_\text{SBR} =1150$~nm. (b) Illustrating power flows for an open-rack and a close-roof mounted PV module: incident solar irradiance $I_m$, electrical output power $P_\text{el}$, reflected power $P_R$, radiative power $P_\text{rad}$ and convective power $P_\text{conv}$. Placing an SBR on top of the PV module increases $P_R$.}
\end{figure}

In this work, we estimate the effect of an \emph{ideal} SBR on the energy yield of a silicon PV module. As shown in Fig.~\ref{fig:ideal}(a), such an ideal SBR would reflect 100\% of the incident light for wavelength longer than an SBR cutoff wavelength ($\lambda>\lambda_\text{SBR}$), which is in the proximity of the bandgap, but would not alter the reflectivity of the PV module for $\lambda<\lambda_\text{SBR}$ (hence, the SBR has 0 reflection and 100\% transmission for $\lambda<\lambda_\text{SBR}$). We further assume that the SBR does not alter the emissive properties of the PV module for $\lambda>4$~\textmu m so that it does not alter radiative cooling. As shown in Fig.\ \ref{fig:ideal}(b), we assess two modules in open-rack and close-roof configurations, respectively.  Further, we present some considerations on the effect of SBRs on the degradation and lifetime of PV modules. All models were implemented using \texttt{Python}, the scripts are shared in the accompanying data publication.

\section*{Results and discussion}

\subsection*{Annual energy yield}

\begin{figure}
    \centering
    \includegraphics{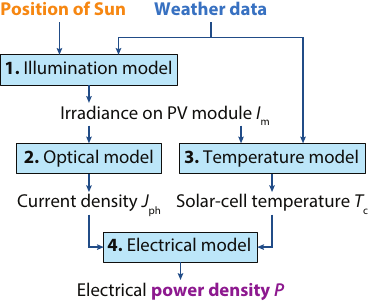}
    \caption{Flowchart illustrating the data sets and models required to calculate the electrical power density and hence energy yield of a PV module.}
    \label{fig:ey-sketch}
\end{figure}

\begin{table*}
\centering
\caption{The six locations at which the model was evaluated, their climate according to the Köppen-Geiger classification,\cite{Peel2007}
%the year for which the weather data was evaluated
and the average ambient temperature $\overline{T}_\text{amb}$ weighted by the global horizontal irradiance (GHI) following the convention of Blom \emph{et al.}\cite{Blom2025} and defined in the ESI, eq.\ (\ref{eq:irr_weighted}). Information on the PV module tilt is given in the ESI, section \ref{sec:tilt}.}
\label{tab:locations}
\begin{tabular}{lcclcc}
\toprule
\textbf{Location} &  \textbf{Latitude} & \textbf{Longitude} & \textbf{Climate} & \textbf{$\overline{T}_\text{amb}$}\\
\midrule
Princeton, New Jersey, USA & 40.33~\textdegree N & \phantom{1}74.66~\textdegree W & Cfa (humid subtropical) & 19.9~\textdegree C\\
Miami, Florida, USA & 25.77~\textdegree N & \phantom{1}80.18~\textdegree W & Am (tropical monsoon) & 27.3~\textdegree C\\
La Paz, Baja California Sur, Mexico & 24.17~\textdegree N & 110.30~\textdegree W & BWh (subtropical hot desert) & 30.1~\textdegree C\\
Golden, Colorado, USA & 39.77~\textdegree N & 105.22~\textdegree W & BSk (cold semi-arid) & 19.6~\textdegree C\\
Ketchikan, Alaska, USA & 55.33~\textdegree N & 131.66~\textdegree W & Cfb (oceanic)  & 13.1~\textdegree C\\
Berlin, Germany & 52.52~\textdegree N & \phantom{1}13.39~\textdegree \makebox[\widthof{W}]{E} & Cfb / Dfb (humid continental) &  19.5~\textdegree C\\
\bottomrule
\end{tabular}
\end{table*}

We calculate the \emph{annual} energy yield (AEY) with a chain of four models as illustrated in Fig.\ \ref{fig:ey-sketch} and described in detail in the \emph{Electronic Supplementary Information} (ESI), section \ref{sec:ey}. These models calculate the output power density $P$ from hourly weather data provided by the National Solar Radiation Database (NSRDB) by the US National Laboratory of the Rockies (NLR), together with the position of the Sun, as follows: first, the \emph{illumination model} calculates the irradiance on the plane of array of the PV module $I_m$. From the irradiance, second, the \emph{optical model} determines the maximum achievable current density $J_\text{ph}$. Using weather data and $I_m$, third, a \emph{temperature model} calculates the solar-cell temperature $T_c$. Fourth, an \emph{electrical model} uses $J_\text{ph}$ and $T_c$ to calculate the electrical output power density $P$. The electrical model contains the temperature coefficients $\beta_V$ and $\beta_J$ for the open-circuit voltage and short-current density, respectively, denoting by which percentage these values change per kelvin increase in temperature. The AEY is then calculated with weather data given for one year. Note that we here calculate the single-module DC AEY and neglect further system-level losses.

\begin{figure}
    \centering
    \includegraphics{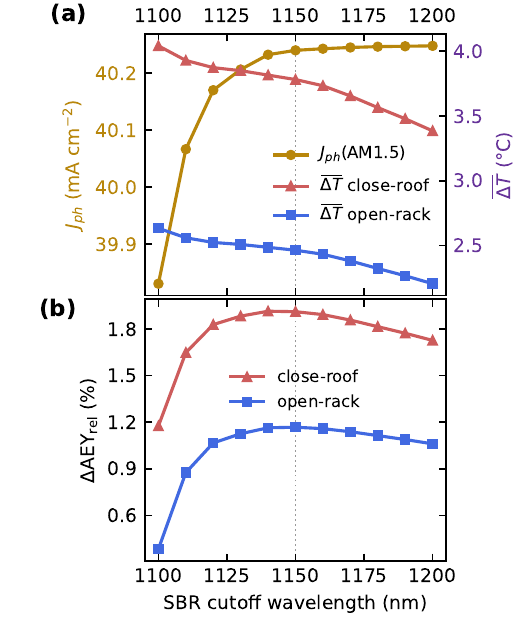}
    \caption{Studying the effect of the sub-bandgap reflector (SBR) cutoff wavelength $\lambda_\text{SBR}$: (a) Maximum achievable current density $J_\text{ph}$ for incident AM1.5 light, and the irradiance-weighted average temperature differences $\overline{\Delta T}=\overline{T_\text{ref}}-\overline{T_\text{SBR}}$ [see ESI, eq.\ (\ref{eq:irr_weighted})] of a solar cell in the PV module without (ref) and with ideal sub-bandgap reflector (SBR). (b) Relative change of the annual energy yield ($\Delta\text{AEY}_\text{rel}=[\text{AEY}_\text{SBR}-\text{AEY}_\text{ref}]/\text{AEY}_\text{ref}$). $J_\text{ph}$ calculated according to ESI, eq.\ (\ref{eq:jsc_am15}), with EQE data from a PERC PV module,\cite{Subedi2020}, see ESI, section \ref{sec:pv-module}. Results shown for open-rack and close roof mounted modules [see Fig.\ \ref{fig:ey-sketch}] in Princeton [see Table \ref{tab:locations}], temperature coefficients of $\beta_V=-0.40~\%\,\text{K}^{-1}$ and $\beta_J=+0.03~\%\,\text{K}^{-1}$.  Temperature data shown for daytime hours, when the Sun is above the horizon.}
    \label{fig:sbr-sweep}
\end{figure}

While the bandgap of silicon is at $\approx 1100$~nm, silicon solar cells usually also generate current for longer wavelength up to $\approx 1200$~nm. Figure \ref{fig:sbr-sweep}(a) shows the maximum achievable current density $J_\text{ph}$ of a PERC solar cell\cite{Subedi2020} as a function of the SBR cutoff wavelength $\lambda_\text{SBR}$, which is considered to be the longest wavelength at which current is generated [see eq. (\ref{eq:jsc_am15})]. For this solar cell, $J_\text{ph}$ rises quickly and saturates at about 1150~nm. Figure \ref{fig:sbr-sweep}(a) also shows the irradiance-weighted average temperature reduction $\overline{\Delta T}$  for open-rack and close roof-mounted PV modules, [see ESI, eq.\ (\ref{eq:irr_weighted})]. $\overline{\Delta T}$ decreases with increasing $\lambda_\text{SBR}$ because of the decreasing fraction of infrared light that is blocked by the SBR. Figure \ref{fig:sbr-sweep}(b) shows the relative change of the annual energy yield $\Delta\text{AEY}_\text{rel}$, which has a broad peak between 1130 and 1160~nm for the considered solar-cell data. For other solar cells with higher EQE values at longer wavelength the maximum may shift to longer wavelength. Therefore, we set $\lambda_\text{SBR}=1150$~nm, the middle of the 1100~nm to 1200~nm region, for the remainder of this work. The results are shown for Princeton, New Jersey, United States.

\begin{table*}[p]
\centering
\caption{Results shown for the six locations with spectrally variable irradiance data $I_\lambda$ for open-rack and close-roof mounted PV panels, with an SBR cutoff wavelength of $\lambda_\text{SBR}=1150$~nm. The table shows the average temperatures $\overline{T_\text{ref}}$ and  $\overline{T_\text{SBR}}$, the maximal $\Delta T_{\max}=\max[T_\text{ref}(t)-T_\text{SBR}(t)]$, and the average temperature differences $\overline{\Delta T}=\overline{T_\text{ref}}-\overline{T_\text{SBR}}$ of a solar cell in the PV module without (ref) and with ideal sub-bandgap reflector (SBR); the annual energy yield without ($\text{AEY}_\text{ref}$) and with ($\text{AEY}_\text{SBR}$) ideal SBR and the relative change of the energy yield ($\Delta\text{AEY}_\text{rel}=[\text{AEY}_\text{SBR}-\text{AEY}_\text{ref}]/\text{AEY}_\text{ref}$). Results are shown for temperature coefficients of $\beta_V=-0.40~\%\,\text{K}^{-1}$ and $\beta_J=+0.03~\%\,\text{K}^{-1}$. Average temperatures were calculated as irradiance-weighted averages, using the plane-of-array irradiance $I_m$ [see ESI, eq.\ (\ref{eq:irr_weighted})]. The values of $\overline{\Delta T}$ and $\Delta$AEY$_\text{rel}$ were calculated before rounding and can therefore deviate from values computed from the rounded numbers shown.}
\label{tab:results_spectral}
\small

\begin{tabular}{l|ccccccc|ccccccc}
\toprule
& \multicolumn{7}{c|}{\textbf{Open-rack mount}} & \multicolumn{7}{c}{\textbf{Close-roof mount}}\\
\textbf{Location} & \textbf{$\overline{T_\text{ref}}$} & \textbf{$\overline{T_\text{SBR}}$} & \textbf{$\Delta T_\text{max}$} & \textbf{$\overline{\Delta T}$} & \textbf{AEY$_\text{ref}$} & \textbf{AEY$_\text{SBR}$} & \textbf{$\Delta$AEY$_\text{rel}$} &  \textbf{$\overline{T_\text{ref}}$} & \textbf{$\overline{T_\text{SBR}}$} & \textbf{$\Delta T_\text{max}$} & \textbf{$\overline{\Delta T}$} & \textbf{AEY$_\text{ref}$} & \textbf{AEY$_\text{SBR}$} & \textbf{$\Delta$AEY$_\text{rel}$} \\
 & \textdegree C & \textdegree C & \textdegree C & \textdegree C & kWh~m$^{-2}$ & kWh~m$^{-2}$ & -- & \textdegree C & \textdegree C & \textdegree C & \textdegree C & kWh~m$^{-2}$ & kWh~m$^{-2}$ & -- \\
\midrule
Princeton & 39.2 & 36.7 & 4.6 & 2.5 & 423 & 428 & 1.2\% & 50.2 & 46.5 & 7.0 & 3.8 & 401 & 409 & 1.9\%\\
Miami & 44.9 & 42.9 & 3.9 & 2.0 & 452 & 457 & 1.0\% & 54.9 & 51.8 & 6.1 & 3.1 & 430 & 437 & 1.6\%\\
La Paz & 52.3 & 49.6 & 4.5 & 2.6 & 548 & 555 & 1.4\% & 64.5 & 60.4 & 6.8 & 4.1 & 513 & 525 & 2.3\%\\
Golden & 39.4 & 36.8 & 4.9 & 2.6 & 490 & 496 & 1.2\% & 51.1 & 47.0 & 7.5 & 4.0 & 462 & 472 & 2.1\%\\
Ketchikan & 29.5 & 27.5 & 4.5 & 2.0 & 272 & 275 & 0.9\% & 38.5 & 35.4 & 6.9 & 3.1 & 261 & 265 & 1.5\%\\
Berlin & 36.0 & 34.0 & 4.2 & 2.1 & 303 & 306 & 1.0\% & 45.5 & 42.3 & 6.4 & 3.2 & 289 & 294 & 1.6\%\\
\bottomrule
\end{tabular}
\end{table*}

\begin{figure*}[p]
    \centering
    \includegraphics[width=0.4\textwidth]{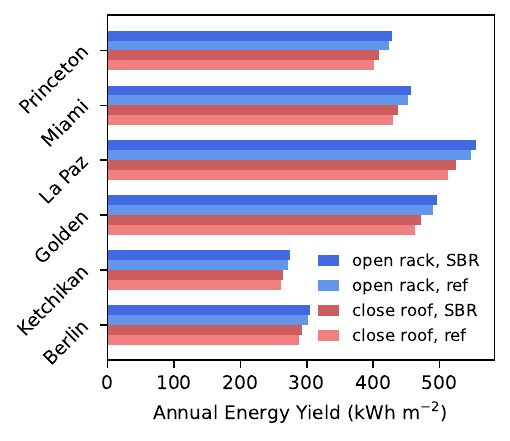}
    \caption{Annual energy yield (AEY) at the six locations for close roof and open rack configurations. The EY values were obtained using  spectrally variable irradiance data and are also shown in Table \ref{tab:results_spectral}.}
    \label{fig:ey_results}
\end{figure*}
\begin{figure*}[p]
    \centering
    \includegraphics[width=0.8\textwidth]{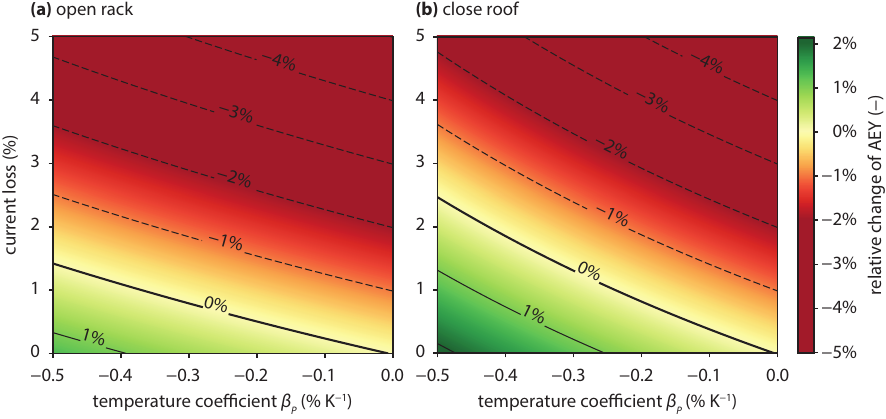}
	\caption{\label{fig:contour}Relative change in annual energy yield $\Delta \mathrm{AEY}_\text{rel}$ as a function of the temperature coefficient for power ($\beta_P$) and the \emph{current loss}, which is the fraction by which a non-ideal SBR reduces $J_\text{sc}$ because of reflection for $\lambda<\lambda_\text{SBR}=1150$~nm. Results shown for Princeton for 2024.}
\end{figure*}

Table \ref{tab:locations} shows geographical and climatic details for Princeton and five other locations with different climates and illumination conditions for which we estimated the effects of ideal sub-bandgap reflectors (SBR) on the AEY. Figure \ref{fig:ey_results} and Table \ref{tab:results_spectral} show the results for the six locations. The average temperature reductions are compatible with previously reported values.\cite{Slauch2019,OrtizLizcano2024} 
Assuming temperature coefficients of $\beta_V=-0.40~\%\,\text{K}^{-1}$ and $\beta_J=+0.03~\%\,\text{K}^{-1}$, an ideal sub-bandgap reflector can increase the annual energy yield between 0.9\% and 1.4\% for a glass/glass module in open-rack mount and between 1.5\% and 2.3\% in close-roof mount. The largest increase is seen in the desert climate of La Paz, while the lowest increase is observed in the  oceanic climate of Ketchikan and the  oceanic to continental climate of Berlin, both with latitudes $>50^\circ$. In the humid climate of Miami, $\Delta\text{AEY}_\text{rel}$ is comparable to Ketchikan and Berlin, even though Miami and La Paz only differ little in latitude. The effect of the location spanning from Alaska to Florida is smaller than the mounting situation of the PV module. This is because, for close-roof mounted modules, convective heat transfer occurs only on the front side of the module, reducing the total convective (non-radiative) heat loss compared to open-rack mounting and therefore increasing the radiative fraction of the heat transfer. Since the sub-bandgap reflector reduces parasitic heating from absorbed sub-bandgap light, its effect on the temperature of close-roof mounted modules is larger than for open-rack mounted modules. Because close-roof mounted modules are hotter than open-rack mounted modules, $\text{AEY}_\text{SBR}(\text{close roof})$ is less than $\text{AEY}_\text{ref}(\text{open rack})$ for the investigated locations.

Now, we extend our analysis to variable temperature coefficients and non-ideal SBRs, where the non-ideality corresponds to a loss in \emph{current density}. The non-ideal SBR might induce some non-intended losses by partially reflecting the incident light for wavelengths shorter than the cutoff wavelength ($\lambda_\text{SBR}=1150$~nm). Figure \ref{fig:contour} shows the relative change in energy yield $\Delta\text{AEY}_\text{rel}$ as a function of the temperature coefficient for power $\beta_P$ and the irradiance loss for open-rack and close-roof mounted modules in Princeton. The results shown in Fig.\ \ref{fig:contour} were obtained with the simple approach [eq. (\ref{eq:p_simple})] explained in the ESI. We see that even small irradiance losses counterbalance the gain in energy yield leading to a net loss. This effect is more severe for open-rack mounted modules.

Until now we showed results for which spectral variations throughout the year were considered. However, this data often is not available, for example in the \emph{typical meteorological year} (TMY) datasets provided by the NSRDB [see ESI, section \ref{sec:nsrdb}]. Table \ref{tab:results_am15} in the ESI shows the results for the same locations as in Table \ref{tab:results_spectral}, but now the spectrum was fixed at AM1.5. We see that in this case the estimated relative change in energy yield differs up to $0.3\%$ from the (more accurate) case where spectral variability is considered.

From these results we can conclude that the effect of SBRs has little to no beneficial effect on the instantaneous power conversion efficiency and the resulting annual energy yield of a single-panel silicon PV system. Another upcoming technology is perovskite photovoltaics with typical bandgaps between 1.2 and 2.3 eV,\cite{yang2018} larger than that of silicon. Quantitatively estimating how an ideal SBR would affect the AEY of a perovskite PV module would require reflectance data up to 2500~nm, which is not available. Therefore, we restrict the following to a qualitative discussion: Because of the higher bandgap, $\lambda_\text{SBR}$ can be shorter for perovskite PV than for silicon PV, which increases the cooling potential achievable by an SBR. However, reported temperature coefficients\cite{Moot2021,jost:2020outdoor} for the power between $\beta_P=-0.08~\%\,\text{K}^{-1}$ and $-0.17~\%\,\text{K}^{-1}$ are significantly below the values for silicon solar cells and reduce the benefit of a passively cooling SBR on the instantaneous power conversion efficiency and the resulting AEY. Nonetheless, SBRs might reduce degradation and hence increase the lifetime of PV modules and the \emph{cumulative} energy yield over the whole module lifetime.

\subsection*{PV module degradation}

Let us now look, how SBRs may affect the degradation of PV modules. Many different mechanisms contribute to the degradation of PV modules. Therefore modelling PV degradation is a complex undertaking, which has been approached with \emph{data-driven} and \emph{analytical} (or \emph{physics-based}) models.\cite{Lindig2018} For assessing, how sub-bandgap reflectors affect the degradation rate, physics-based models, which have the module temperature as input, are required. Temperature is well known to affect the reaction rate of chemical reactions, and therefore the degradation rate of devices such as PV modules. For polymers, which are present in all PV modules, an often used \emph{rule of thumb} states that the reaction rate doubles with every 10~\textdegree C increase in temperature.\cite{Hukins2008} 

For silicon PV modules, several authors proposed physical models to predict the degradation rate based on external parameters. Bala Subramaniyan and coworkers propose a formula with an Arrhenius term and additional coefficients to account for UV dose, daily temperature differences and humidity.\cite{BalaSubramaniyan2018} Kaaya and coworkers combine three Arrhenius terms for UV-based, humidity-based and temperature-cycle-based degradation.\cite{Kaaya2019,Kaaya2021,Kaaya2024} Blom \emph{et al}.\ use the same degradation mechanisms and add a term for light-induced degradation (LID).\cite{Blom2026} Besides LID, all mechanisms are described with Arrhenius terms and additional factors, but the different publications use significantly varying activation energies for the same mechanisms. Adding further complexity, recent research showed that a small number of PV modules degrades much faster than average.\cite{Tang2026}

Because of the complexity summarized above, our aim in this work is to obtain a basic, qualitative understanding of how the temperature reduction caused by an SBR could affect the degradation of PV modules. Because an SBR would neither affect the UV dose received by the module, the humidity of the air, nor light-induced degradation, which depends on cumulative illumination rather than temperature, we therefore restrict ourselves to a simple \emph{Arrhenius} model, where the degradation rate $k$ is given by
\begin{equation*}
    k = \gamma \cdot\exp\left(-\frac{E_a}{k_BT}\right),
\end{equation*}
with the activation energy $E_a$, the temperature $T$ (in kelvin), the Boltzmann constant $k_B$ and a pre-exponential factor $\gamma$.
While an SBR might also reduce $\Delta T$ relevant to thermal-cycling-induced degradation, quantifying this effect goes beyond the scope of this qualitative assessment.
The models mentioned above use Arrhenius terms for the different degradation mechanisms with various activation energies ranging from $E_a=0.12$~eV to $\approx 0.89$~eV.\cite{BalaSubramaniyan2018,Kaaya2019,Kaaya2021,Jiang2023,Kaaya2024,Blom2026} For this qualitative assessment we consider activation energies ranging from 0.4~eV to 0.8~eV. Further, here we use non-spectral weather data for a \emph{typical meteorological year} (TMY, see ESI, Section \ref{sec:nsrdb}), because for the long-term degradation the representativeness of the TMY temperature data outweighs the missing spectral information. More details on the implementation of our degradation model are given in the ESI, section \ref{sec:degradation}.

\begin{figure}
    \centering
    \includegraphics[width=\linewidth]{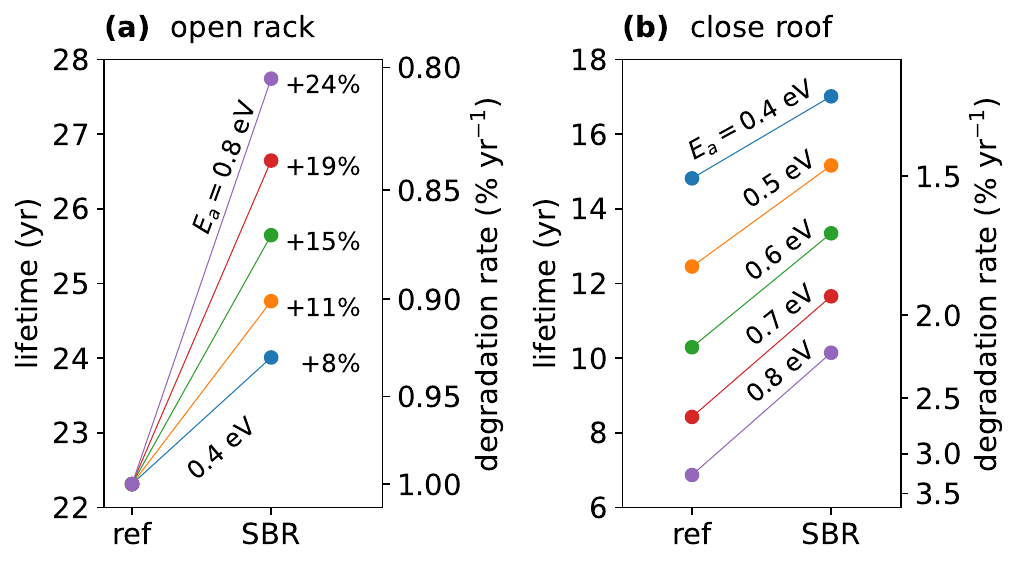}
    \caption{The effect of SBR on the lifetime $t_\text{80\%}$ and the annual degradation rate $k_\text{yr}$ of PV modules mounted in (a) open-rack and (b) close roof condition in Princeton. These results were obtained using eq.\ (\ref{eq:k_yr}), in which the pre-exponential factor $\gamma$ was determined such that the degradation rate for the reference module in open-rack mount is $k_\text{yr} = 1.0~\%~\text{yr}^{-1}$ for all investigated activation energies. The percentage in (a) shows the relative increase of lifetime for a PV module with SBR w.r.t. a reference module. Results obtained with non-spectral \emph{typical meteorological year} (TMY) data.}
    \label{fig:degradation}
\end{figure}

Figure~\ref{fig:degradation} quantifies the impact of the SBR on module lifetime by linking the temperature reduction to a thermally-activated degradation process, for activation energies between 0.4 and 0.8~eV. For this analysis, we set the degradation rate of a reference PV module mounted in open-rack configuration to $k_\text{yr}=1~\%~\text{yr}^{-1}$ for all activation energies, corresponding to a lifetime of about 20~years, and determined $\gamma$ in eq.~(\ref{eq:k_yr}) accordingly. For the open-rack configuration, shown in Fig.~\ref{fig:degradation}(a), an ideal SBR increases lifetime by 8\% for $E_a=0.4$~eV to 24\% for $E_a=0.8$~eV.
For close-roof mount, shown in Fig.~\ref{fig:degradation}(b), reference-module lifetimes are much shorter and, unlike for open-rack mount, vary substantially with $E_a$, because $\gamma$ is calibrated to a fixed open-rack reference lifetime. Qualitatively, the SBR's relative benefit is larger for close-roof mount, since degradation is more strongly thermally activated at its higher operating temperature. On a sunny summer day, the temperature difference between the two mounting configurations approaches 20~\textdegree C, as shown in Fig.~\ref{fig:temp_princeton} in the ESI. Even with an ideal SBR, the lifetime of close-roof mounted modules remains shorter than that of the open-rack reference module (i.e., without SBR), for all activation energies considered.

\begin{figure*}
    \centering
    \includegraphics{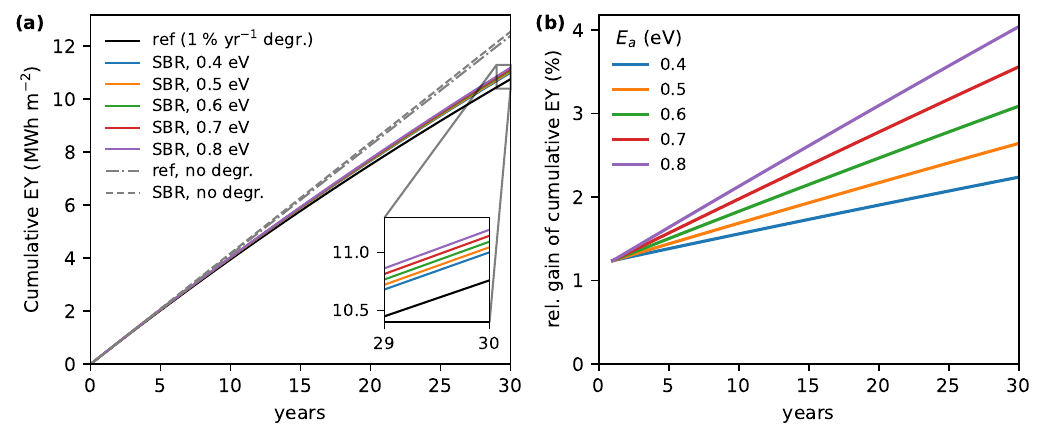}
    \caption{(a) The cumulative energy yield $\Sigma$EY for an \emph{open rack} mounted PV module in Princeton, as defined in eq.~(\ref{eq:ey_cum}) in the ESI. Data is shown for a reference module with no SBR and an estimated annual degradation rate of 1~\%~yr$^{-1}$ and of PV modules with an ideal SBR for different activation energies. Further, the dashed and dot-dashed lines show the cumulative energy yield for a PV module with and without an ideal SBR, when no degradation is assumed. (b) The relative gain in cumulative energy yield $\Gamma$, calculated with $\Gamma = (\Sigma\text{EY}_\text{SBR}-\Sigma\text{EY}_\text{ref})/\Sigma\text{EY}_\text{ref}$, for different activation energies. Results obtained with non-spectral \emph{typical meteorological year} (TMY) data.}
    \label{fig:degradation2}
\end{figure*}

Figure \ref{fig:degradation2}(a) shows the \emph{cumulative} energy yield ($\Sigma$EY) for an \emph{open rack} mounted PV module in Princeton considering degradation, as defined in eq.~(\ref{eq:ey_cum}) in the ESI. Data is shown for a reference module with no SBR and an estimated annual degradation rate of $k_\text{yr} = 1~\%\,\text{yr}^{-1}$ and of PV modules with an ideal SBR for different activation energies. The gray dashed lines show $\Sigma$EY without degradation.

Figure \ref{fig:degradation2}(b) shows the \emph{relative gain in cumulative energy yield}, calculated via
\begin{equation*}
    \Gamma(T) = \frac{\Sigma\text{EY}_\text{SBR}(T)-\Sigma\text{EY}_\text{ref}(T)}{\Sigma\text{EY}_\text{ref}(T)}.
\end{equation*}
After the first year, $\Gamma(0)=1.2\%$. Over the years, $\Gamma(T)$ increases because the modules with SBR degrade slower than the reference module. The effect is more pronounced when the activation energy is higher. After 30 years, our simple approach gives values between $\Gamma(30\text{~yrs})=2.2\%$ and $4.0\%$ for activation energies between 0.4~eV and 0.8~eV.

Even though this qualitative analysis does not allow us to predict how an SBR would affect the degradation rate of a specific PV module, it demonstrates the potential benefit of SBRs for the longevity and stability of PV modules---an effect that might be even more relevant for perovskite solar cells, where temperature variations have been identified as critical for long-term stability.\cite{Wu2025} Together with the modest AEY gains discussed above, these results show that even where the annual energy gain of a non-ideal SBR is modest or even negative, the associated reduction in operating temperature can still provide a meaningful benefit in module lifetime and cumulative energy yield.

\subsection*{Sensitivity Analysis and Limitations}

Having presented results for AEY and module degradation, we now examine how sensitive these conclusions are to key parameters and we discuss limitations of our modelling approach.

\begin{description}
    \item[Temperature coefficient and non-ideal SBR losses] Revisiting Fig.~\ref{fig:contour}, this figure can be read as sensitivity analysis showing both the effects of varying the temperature coefficient for power $\beta_P$ and deviating from an ideal SBR. We see that the potential gains in EY become smaller with decreasing $\beta_P$ and that even small current losses because of a non-ideal SBR may turn the effect of an SBR from a net gain to a net loss. For $\beta_P=-0.3~\%\,\text{K}^{-1}$ a current loss of about 1\% would cancel the gain caused by temperature reduction for an open-rack mounted module.

    \item[SBR cutoff wavelength] As discussed above and shown in Fig.~\ref{fig:sbr-sweep}, the optimal cutoff wavelength lies broadly between 1130 and 1160~nm irrespective of the mounting configurations. We therefore use $\lambda_\text{SBR}=1150$~nm, the midpoint of the 1100--1200~nm transition region, throughout this work, rather than the device-specific optimum.

    \item[Interannual variability] The results presented above use a single year of weather data per location. Table \ref{tab:results_annual} in the ESI shows results for five consecutive years (2020--2024) for Princeton. We find that while absolute AEY varies by up to 8\% between years, $\Delta\text{AEY}_\text{rel}$ remains stable, ranging from 1.1\% to 1.2\% for open-rack and from 1.9\% to 2.0\% for close-roof mounting. Because both reference and SBR modules experience the same year-to-year variability, much of these variations cancels in the relative comparison.

    \item[Activation energy for degradation] The benefit of an SBR on module degradation depends strongly on the assumed activation energy $E_a$. As shown in Fig.~\ref{fig:degradation}, for an open-rack module, lifetime increases by 8\% to 24\% over the literature range $E_a=0.4$--$0.8$~eV. Fig.~\ref{fig:degradation2} shows that, for the same configuration, the cumulative lifetime energy yield increases by 2.2\% to 4.0\%. These ranges represent our estimate of the uncertainty in the degradation-related benefit of an SBR.
\end{description}

\subsubsection*{Limitations}
\begin{itemize}
    \item We consider an \emph{ideal} SBR with a perfectly sharp reflectivity step at $\lambda_\text{SBR}$. A real coating would introduce optical losses, for example due to dispersion, parasitic absorption, or interference effects, which would reduce the current density. Fig.~\ref{fig:contour} shows that even modest non-ideal losses substantially reduce the benefit.
    \item We used reflectance and EQE data for one PERC device. While we expect our qualitative conclusions to hold for silicon PV modules in general, they do not quantitatively transfer to other technologies with a different bandgap, such as perovskite photovoltaics.
    \item We calculated module and cell temperatures with the empirical \emph{Sandia Module and Cell Temperature Model}, using the parameters for glass/glass modules in open-rack and close-roof mount (ESI, Section~\ref{sec:temperature}). The SBR is represented by a reduced irradiance input, with unchanged empirical coefficients, and the model cannot easily be adapted to other module types. A more general analysis would require a physical heat transfer energy balance model, such as the Fuentes model,\cite{fuentes:1987} extended with spectrally resolved emissivity and absorptivity.
    \item Our degradation model uses a single effective Arrhenius term to isolate the thermally-activated, SBR-addressable degradation channel; it excludes swing-driven (thermal cycling) and light-induced degradation, and likely overestimates the fraction of total degradation an SBR can mitigate.
    \item We further neglect temperature-dependent bandgap narrowing and the angle-of-incidence dependence of reflectivity; the latter affects the reference and SBR modules similarly and should largely cancel in $\Delta$AEY$_\text{rel}$.
\end{itemize}

Despite these idealizations, the moderate AEY benefit and the more substantial degradation-related benefit of an SBR are consistent across the range of parameters and assumptions examined in this manuscript.

\section*{Conclusions and Outlook}
We estimated how an idealized passive cooling sub-bandgap reflector (SBR) would affect the annual and cumulative energy yield of silicon PV modules for six locations in North America and Europe. For a PV module with temperature coefficients for voltage and current of $\beta_V=-0.40~\%\,\text{K}^{-1}$ and $\beta_J=+0.03~\%\,\text{K}^{-1}$, respectively, an ideal SBR would moderately increase the annual energy yield (AEY) between 0.9\% and 1.4\% for open-rack mounted modules and between 1.5\% and 2.3\% for close-roof mounted PV modules. This small increase can be rapidly negated in case of a non-ideal SBR with reduced transmissivity in the spectral solar cell absorption regime.  Whether a non‑ideal SBR provides a benefit in actual deployments strongly depends on the location and the optical properties of the coating for short wavelengths, where the incident light can be utilized by the solar cells.

Beyond effects on the instantaneous power conversion efficiency and hence the AEY, a SBR might reduce the degradation of the PV module because of the reduced temperature. When describing degradation by a simple Arrhenius approach, an ideal SBR would increase the cumulative energy yield over 30 years between 2.2\% and 4.0\% for typical activation energies between 0.4~eV and 0.8~eV for an open-rack mounted PV module in Princeton, New Jersey, USA.

From this work arise several interesting and relevant topics for future research. Amongst these are the design and fabrication of SBRs with production costs that justify the modest expected gains in efficiency and lifetime. Further, here we estimated the effect of an SBR on the PV module temperature assuming a single isolated PV module. It should be investigated how thermal management with SBRs affects the surrounding temperatures in large PV plants with many PV modules and in urban environments.

\section*{Author contributions}
\textbf{Klaus Jäger}: Conceptualization, methodology, software, investigation, data curation, writing - original draft, visualization, funding acquisition. \textbf{Jyotirmoy Mandal}: conceptualization, writing - review \& editing. \textbf{Barry P.\ Rand}: conceptualization, writing - review \& editing. \textbf{Forrest Meggers}: conceptualization, writing - review \& editing. \textbf{Christiane Becker}: conceptualization, investigation, writing - review \& editing.

\section*{Conflicts of interest}
There are no conflicts to declare.

\section*{Data availability}

The code used for generating the results published in this manuscript, together with examples, is archived on Zenodo as version~0.2.0 of the Python package \texttt{pv\_tandem}: \href{https://doi.org/10.5281/zenodo.24114973}{\texttt{https://doi.org/10.5281/zenodo.24114973}}. The code can be found in the file \texttt{pv\_tandem/thermal.py}; the directory \texttt{examples/thermal} contains two simple examples and \texttt{Jupyter} notebooks for generating Figs.~\ref{fig:sbr-sweep}, \ref{fig:contour}, \ref{fig:degradation} and \ref{fig:degradation2}. The package is developed at \href{https://github.com/HZBSolarOptics/pv_tandem/}{\texttt{https://github.com/HZBSolarOptics/pv\_tandem/}}. The input data are weather and irradiance data from the National Solar Radiation Database (NSRDB)\cite{Sengupta2018}, which are included in the archive, and reflectance and quantum efficiency data of the PERC module, which were digitized from Figs.~5 and~6 of Subedi \emph{et al.}\cite{Subedi2020}

\section*{Acknowledgements}

KJ thanks Princeton University to make this work possible by allowing him to stay at the Andlinger Center for Energy and the Environment as a Gerhard R.\ Andlinger Visiting Fellow. 
Parts of the research were performed at the Berlin Joint Lab for Optical Simulations for Energy Research (BerOSE) of Helmholtz-Zentrum Berlin für Materialien und Energie, Zuse Institute Berlin, and Freie Universität Berlin. 

During the preparation of the original submission, KJ used Microsoft Copilot (version unknown) to prepare the code for publication on GitHub and to search the literature. During the revision of this manuscript, KJ used Claude Sonnet (v. 5 and 5.5) and Claude Opus (v. 5 and 5.5) from Anthropic via \href{https://claude.ai}{\texttt{https://claude.ai}} to draft and revise text passages, to review and refactor the Python code used for the energy-yield simulations, to write plotting scripts for figures, to check tabulated values for consistency, and to search the literature. Prompts consisted of the relevant manuscript text, code or data together with a description of the task; transcripts are available on request. The authors have reviewed and edited the output and take full responsibility for the content of this publication.

% -------------------------------------------------
% Electronic Supplementary Information
% -------------------------------------------------
\clearpage
\begin{widetext}
\begin{center}
%\centering
\large
\textbf{Electronic Supplementary Information}
\end{center}
\end{widetext}

\beginsupplement

\section{Energy yield model}
\label{sec:ey}

In this section we describe, how the annual energy yield (AEY) is calculated.
Figure \ref{fig:ey-sketch} in the main manuscript illustrates the data sets and models that need to be combined to estimate the power output of a PV module and subsequently its annual energy yield (AEY). The data sources used as input for the model are described in section~\ref{sec:input}.

\subsection{Illumination model}
\label{sec:illumination}
We use a view-factor based \textbf{illumination model} to calculate the total irradiance $I_m(t)$ onto the solar module based on direct normal irradiance (DNI) and diffuse horizontal irradiance (DHI) data [see section \ref{sec:nsrdb}], the position of the Sun and geometrical factors.\cite{jaeger:2020oe} We used a module height above the ground of 1.5~m, a module width of 1.1~m and a module length of 1.7~m. Further, we assumed stand-alone modules, hence no shadowing effect, which we implemented by setting the distance between two rows of modules to 100~m. The albedo of the ground was set to 30\%. For each location, we optimized the module tilt to maximize the annual radiant exposure, as described below.

\subsection{Module tilt}
\label{sec:tilt}

We set the module tilt $\theta_\text{mod}$ such that the radiant exposure (the incident irradiance onto the module $I_m$ integrated over a full year) on the front of the PV module is maximized, 
\begin{equation}
    \sum_\text{year} I_m(\theta,t)\Delta t \rightarrow \max.
\end{equation}
$I_m$ is calculated with the view-factor model described in section \ref{sec:illumination}. The optimization is done with the function \texttt{minimize\_scalar} provided by \texttt{SciPy}.\cite{2020SciPy-NMeth}

Table \ref{tab:tilt} shows the optimal tilt $\theta_\text{opt}$ for the six locations. As input for the tilt optimization, we used typical meteorological year (TMY) data \cite{wilcox:2008} [see section \ref{sec:nsrdb}]. Based on this angle we downloaded a spectral annual dataset from the NSRDB database. Since their web interface only allows for integer values as tilt angles, we used the closest integer $\theta_\text{spec}$.

In summary, for calculating the energy yield in this work, we set the module tilt to $\theta_\text{mod} = \theta_\text{opt}$ and used spectral input data for $\theta_\text{spec}$.

\begin{table}
\centering
\caption{The tilt angles of the PV modules for the six locations. First, the optimal tilt angle $\theta_\text{opt}$ was determined for TMY datasets. For the spectral datasets the tilt angle with the closest integer value $\theta_\text{spec}$ was downloaded. The years denote the years of the TMY and spectral datasets as in the NSRDB.}
\label{tab:tilt}

\begin{tabular}{lcc|cc|cc}
\toprule
\textbf{Location} &  \textbf{Latitude} & \textbf{Longitude} & \multicolumn{2}{c|}{\textbf{TMY data}} & \multicolumn{2}{c}{\textbf{Spec. data}}\\
&&& year & $\theta_\text{opt}$ & year & $\theta_\text{spec}$ \\
\midrule
Princeton & 40.33~\textdegree N & \phantom{1}74.66~\textdegree W & 2024 & 34.5\textdegree & 2024 & 35\textdegree\\
Miami & 25.77~\textdegree N & \phantom{1}80.18~\textdegree W & 2022 & 23.0\textdegree & 2024 & 23\textdegree\\
La Paz & 24.17~\textdegree N & 110.30~\textdegree W & 2022 & 22.3\textdegree & 2024 & 22\textdegree\\
Golden & 39.77~\textdegree N & 105.22~\textdegree W & 2022 & 36.9\textdegree & 2024 & 37\textdegree\\
Ketchikan & 55.33~\textdegree N & 131.66~\textdegree W & 2022 & 37.6\textdegree & 2024 & 38\textdegree\\
Berlin & 52.52~\textdegree N & \phantom{1}13.39~\textdegree \makebox[\widthof{W}]{E} & 2022 & 32.1\textdegree & 2022 & 32\textdegree\\
\bottomrule
\end{tabular}
\end{table}

\subsection{Optical model}
\textbf{Optical models} are used to estimate the maximum achievable current density $J_\text{ph}$ generated by the solar module. Instead of involved optical modelling, we estimate $J_\text{ph}$ via the external quantum efficiency (EQE) of a PV module [see section \ref{sec:pv-module}] with
\begin{equation}
    \label{eq:jsc_spectral}
    J_\text{ph}(t) = \frac{e}{hc}\,\frac{I_m(t)}{I^\text{NSRDB}(t)}
    \int_{\lambda_\text{min}}^{\lambda_\text{co}} \lambda\,\mathrm{EQE}(\lambda)\, I^\text{NSRDB}_\lambda(t)\,\dif\lambda,
\end{equation}
where $e$ is the elementary charge, $h$ is Planck's constant, $c$ is the speed of light \emph{in vacuo}, $I^\text{NSRDB}_\lambda$ is the spectral irradiance for a given location and time [see \ref{sec:nsrdb}], and
\begin{equation*}
    \label{eq:itot}
    I^\text{NSRDB}(t) = \int_{\lambda_\text{min}}^{\lambda_\text{max}} I^\text{NSRDB}_\lambda(t)\,\dif\lambda.
\end{equation*}
The ratio $I_m/I^\text{NSRDB}$ is to correct for the illumination on the PV module as described in Section \ref{sec:illumination}. If no spectral data is available, the AM1.5 spectrum is used instead,
\begin{equation}
    \label{eq:jsc_am15}
    J_\text{ph}(t) = \frac{e}{hc}\,\frac{I_m(t)}{I^\text{AM1.5}}
    \int_{\lambda_\text{min}}^{\lambda_\text{co}} \lambda\,\mathrm{EQE}(\lambda)\, I^\text{AM1.5}_\lambda\,\dif\lambda
\end{equation}
with $I^\text{AM1.5} = 1000$~W~m$^{-2}$. Here, the integral only needs to be evaluated once.

The integration borders are $\lambda_\text{min} = 280$~nm and $\lambda_\text{max} = 4000$~nm. The \emph{cutoff wavelength} $\lambda_\text{co}$ depends on whether $J_\text{ph}$ is evaluated for a reference module or a module with SBR: $\lambda_\text{co,ref} = 1200$~nm and $\lambda_\text{co,SBR} = \lambda_\text{SBR}\in[1100\,\text{nm},\,1200\,\text{nm}]$.

We did not account for temperature-dependent variations of the optical properties, such as bandgap narrowing, and therefore temperature-dependent variations of the EQE.

\subsection{Temperature model}
\label{sec:temperature}
\begin{table}
    \caption{Parameters used in the Sandia Module and Cell Temperature Models for glass/glass modules which are mounted in open-rack and close-roof configuration,\cite{king2004} as illustrated in Fig.~\ref{fig:ideal}(b).}
    \label{tab:sapm}
    \centering
    \begin{tabular}{cccc}\toprule
       Mount  & $a$ (--) & $b$ (s~m$^{-1}$) & $\Delta T$ (\textdegree C)  \\ \midrule
       open rack   & --3.47 & --0.0594  & 3 \\
       close roof  & --2.98 & --0.0471  & 1 \\ \bottomrule
    \end{tabular}
\end{table}

Several \textbf{temperature models} are available to estimate the temperatures of the PV module and the solar cell as a function of irradiance, ambient temperature and wind speed. We used the empirical \emph{Sandia Module and Cell Temperature Models},\cite{king2004} because they offer parameters for different module configurations. The module temperature $T_m$ and cell temperature $T_c$ are given by
\begin{align}
    T_m(t) &= I_m(t)\cdot\exp\left[a+b\cdot v_W(t)\right]+T_a(t),\\
    T_c(t) &= T_m(t) + \frac{I_m(t)}{I_m(\mathrm{AM1.5})}\Delta T,
\end{align}
with the parameters $a$, $b$ and $\Delta T$. In this work we look at glass/glass modules in open-rack and close-roof mount configurations, as illustrated in Fig.~\ref{fig:ideal}(b). The parameters for these module types are given in Table \ref{tab:sapm}.

Throughout this work, we report \emph{irradiance-weighted temperatures},\cite{Blom2025} defined for a time-dependent temperature dataset $T(t)$ as
\begin{equation}
    \overline{T} = \frac{\sum_t I(t)\,T(t)}{\sum_t I(t)},
    \label{eq:irr_weighted}
\end{equation}
where $I(t)$ is the global horizontal irradiance (GHI) for the ambient temperature in Table~\ref{tab:locations}, and the plane-of-array irradiance $I_m(t)$ for the module and cell temperatures reported elsewhere in this work (e.g.\ $\overline{T}_m$, $\overline{T}_c$, $\overline{\Delta T}$). This weighting emphasizes conditions during which the module is most productive and inherently excludes hours with zero irradiance.

\begin{figure}
    \centering
    \includegraphics[width=\linewidth]{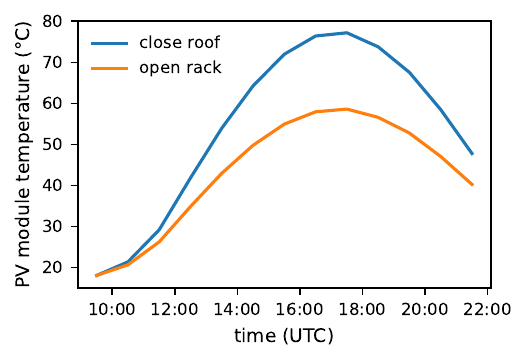}
    \caption{The temperature of a PV module (with no SBR) in open-rack and close-roof configurations in Princeton on 10 July 2005. Times are given in UTC. The maximum temperature difference between the two configurations is 18.6~\textdegree C.}
    \label{fig:temp_princeton}
\end{figure}

Figure \ref{fig:temp_princeton} shows the temperature of a PV module mounted in \emph{open rack} and \emph{close roof} conditions in Princeton on 10 July 2005. We see that the mounting type can lead to large differences in temperature---up to 18.6~K for the shown example.

\subsection{Temperature effect of ideal sub-bandgap reflector (SBR)}
To estimate the effect of an ideal sub-bandgap reflector (SBR), we assumed that the irradiance $I_m$ is reduced by the fraction that would be reflected by an ideal SBR,
\begin{align}
    I_m(t) \rightarrow I_m^\mathrm{SBR}(t) &= I_m(t)\cdot \left[1-f_\text{IR}(t)\right],\\
    \label{eq:f_ir}
    f_\text{IR}(t) &= \frac{\int_{\lambda_\text{SBR}}^{\lambda_{\text{max },R}}(1-R_\lambda)I_\lambda(t)\,\dif\lambda}{\int_{\lambda_\text{min}}^{\lambda_\text{max}}I_\lambda(t)\,\dif\lambda},
\end{align}
where $f_\text{IR}(t)$ is the fraction of the irradiance for wavelength longer than the SBR cutoff wavelength $(\lambda>\lambda_{\text{SBR}})$ that would be absorbed by the reference PV module without SBR. The spectral irradiance $I_\lambda$ is equal to $I^\text{NSRDB}_\lambda$ if spectral irradiance data is available [see section \ref{sec:nsrdb}], and $I^\text{AM1.5}_\lambda$ else. $R_\lambda$ is the spectral reflectivity of the PV module. Here we use data for a PERC PV module as described in section \ref{sec:pv-module}. $R_\lambda$ data is provided between 300~nm and 2500~nm, and therefore $\lambda_{\text{max, }R} = 2500$~nm. Hence, we assume that our SBR does not affect the interval [2500\,nm,~4000\,nm], which contains about 0.8\% of the total irradiance for AM1.5.  The integration borders of the integral in the denominator of eq.\ (\ref{eq:f_ir}) are $\lambda_\text{min} = 280$~nm and $\lambda_\text{max} = 4000$~nm. 

Details on the spectral irradiance data $I^\text{NSRDB}_\lambda$ are given in section \ref{sec:nsrdb}. If no spectral data is given, AM1.5 data can be used.  For AM1.5 and $\lambda_\text{SBR}=1150$~nm we have  $f_\text{IR}(\mathrm{AM1.5})=$12.9\% and hence $I_m^\text{SBR}=87.1\%\cdot I_m$.

\begin{figure*}
    \centering
    \includegraphics{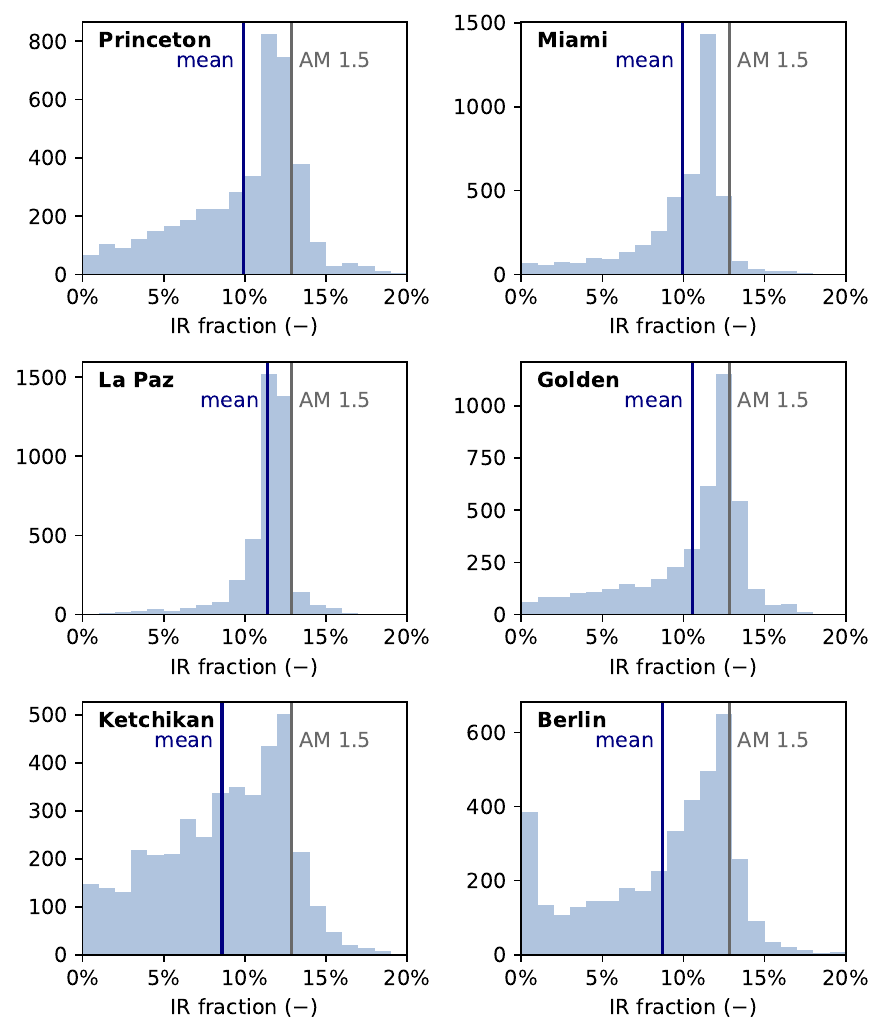}
    \caption{Histograms of the infrared irradiance fraction $f_\text{IR}$ as defined in Eq.~(\ref{eq:f_ir}) for an SBR cutoff wavelength of $\lambda_\text{SBR}=1150$~nm for the six locations. Data is shown for the annual datasets with the tilt angles as specified in Table~\ref{tab:tilt}. Further, the mean infrared fraction $\overline{f_\text{IR}}$ and $f_\text{IR}(\text{AM1.5})= $12.9\% are shown. Note that the histogram only accounts for data where the total irradiance is larger than 0, $I_m>0$.}
    \label{fig:histograms}
\end{figure*}

Figure \ref{fig:histograms} shows histograms for $f_\text{IR}$ for the annual datasets for the six locations for $\lambda_\text{SBR}=1150$~nm. We see that $f_\text{IR}$ is smaller than $f_\text{IR}(\mathrm{AM1.5})$ for most datapoints and therefore the mean $\overline{f_\text{IR}}<f_\text{IR}(\mathrm{AM1.5})$ for all six locations. We further note that $\overline{f_\text{IR}}$ is smallest for the two northern locations Ketchikan and Berlin. It is highest for La Paz with a desert climate, where the irradiation may be closer to AM1.5 because of little cloud coverage.

\subsection{Electrical Model}
Last, we need an \textbf{electrical model} to estimate the generated electric power density $P(t)$. Here, we used a one-diode model to calculate the current-voltage characteristics from which the maximum power point is determined numerically. This model contains the temperature coefficients of the solar module for open-circuit voltage ($\beta_V$) and short-circuit current density ($\beta_J$).\cite{Tillmann2022}

In a simpler but quicker approach that directly incorporates the temperature coefficient for power $\beta_P$, the output power density is estimated with

\begin{equation}
    \label{eq:p_simple}
    P(t)=\eta I^\text{AM1.5}\frac{J_\text{ph}(t)(1-\ell)}{J_\text{ph}^\text{AM1.5}}\times\left[1+\beta_P\left(T_c(t)-25^\circ\text{C}\right)\right].
\end{equation}
Here, $\eta$ is the power conversion efficiency of the PV module at STC conditions, $I^\text{AM1.5}$ is the AM1.5 irradiance, $J_\text{ph}$ is the current density according to eq.\ (\ref{eq:jsc_spectral}), $J_\text{ph}^\text{AM1.5}$ is the current density under AM1.5 illumination (calculated using eq.\ \ref{eq:jsc_am15} with $I_m = I^\text{AM1.5}$), and $\ell$ is the \emph{current loss} defined as the fraction by which the current density is reduced because of a non-ideal SBR. 

When no spectral NSRDB data is provided, equation (\ref{eq:p_simple}) simplifies to
\begin{equation}
    P(t)=\eta I_m(t)(1-\ell)\left[1+\beta_P\left(T_c(t)-25^\circ\text{C}\right)\right].
\end{equation}

\subsection{Annual energy yield}
The \textbf{annual energy yield} is estimated with
\begin{equation}
    \mathrm{AEY} = \sum_\text{year} P(t)\Delta t,
\end{equation}
where the sum extends over all times when the Sun is above the horizon.

\section{Model input}
\label{sec:input}

\subsection{Weather data}
\label{sec:nsrdb}
We used hourly \textbf{weather data} available in the National Solar Radiation Database (NSRDB) by the US National Laboratory of the Rockies (NLR).\cite{Sengupta2018} These data include, amongst others, ambient temperature $T_a$, wind speed $v_w$, and several irradiance parameters: global horizontal irradiance (GHI), direct normal irradiance (DNI) and diffuse horizontal irradiance (DHI). The NSRDB provided typical meteorological year (TMY) data \cite{wilcox:2008} because it better represents the climate of a location than an annual dataset. In addition, the NSRDB provides annual datasets that include the spectral irradiance $I_\lambda(t)$ on a pre-defined module plane, where the subscript $\lambda$ denotes that $I_\lambda$ is given as a function of the wavelength. Because spectral information is important for assessing the effect of an SBR, we use these annual data sets instead of TMY data that do not contain this spectral information.

\subsection{PV module data}
\label{sec:pv-module}

We use reflectivity $R$ and external quantum efficiency EQE data for a typical PV module with passivated emitter and rear contact (PERC) solar cells by Subedi \emph{et al.}\cite{Subedi2020} We selected this publication because it presents reflectance data up to 2500~nm, which is essential for assessing the effect of sub-bandgap reflectors. Most publications only provide reflectance data up to about 1200~nm. From the publication, we use the simulated reflection data $R_\text{enc}$ for the encapsulated PERC B module shown in Fig. 5b (data with scattering included). For the external quantum efficiency, only data for the bare cell, EQE$_\text{bare}$, is given in Fig 6 (we use the simulated dataset for PERC B). From this, we estimate the EQE$_\text{enc}$ for an encapsulated module with
\begin{equation}
	\text{EQE}_\text{enc}(\lambda) = \text{EQE}_\text{bare}(\lambda)\frac{1-R_\text{enc}(\lambda)}{1-R_\text{bare}(\lambda)},
\end{equation}
where $R_\text{bare}$ is the simulated data (with scattering) from Fig 5a. EQE$_\text{enc}$ is used as input for eqs.\ (\ref{eq:jsc_spectral}) and (\ref{eq:jsc_am15}).

\begin{table*}
\caption{Results shown for the six locations with the same annual dataset as in Table \ref{tab:results_spectral} but a fixed AM1.5 spectrum for open-rack and close-roof mounted PV panels, with an SBR cutoff wavelength of $\lambda_\text{SBR}=1150$~nm. The table shows the average temperatures $\overline{T_\text{ref}}$ and  $\overline{T_\text{SBR}}$, the maximal $\Delta T_{\max}=\max[T_\text{ref}(t)-T_\text{SBR}(t)]$, and the average temperature differences $\overline{\Delta T}=\overline{T_\text{ref}}-\overline{T_\text{SBR}}$ of a solar cell in the PV module without (ref) and with ideal sub-bandgap reflector (SBR); the annual energy yield without ($\text{AEY}_\text{ref}$) and with ($\text{AEY}_\text{SBR}$) ideal SBR and the relative change of the energy yield ($\Delta\text{AEY}_\text{rel}=[\text{AEY}_\text{SBR}-\text{AEY}_\text{ref}]/\text{AEY}_\text{ref}$). Results are shown for temperature coefficients of $\beta_V=-0.40\,\%\,\text{K}^{-1}$ and $\beta_J=+0.03\,\%\,\text{K}^{-1}$. Average temperatures were calculated as irradiance-weighted averages, using the plane-of-array irradiance $I_m$ [see eq.\ (\ref{eq:irr_weighted})]. The values of $\overline{\Delta T}$ and $\Delta$AEY$_\text{rel}$ were calculated before rounding and can therefore deviate from values computed from the rounded numbers shown.}
\label{tab:results_am15}
\centering
\begin{tabular}{l|ccccccc}
\toprule
\textbf{Location} & \textbf{$\overline{T_\text{ref}}$} & \textbf{$\overline{T_\text{SBR}}$} & \textbf{$\Delta T_\text{max}$} & \textbf{$\overline{\Delta T}$} & \textbf{AEY$_\text{ref}$} & \textbf{AEY$_\text{SBR}$} & \textbf{$\Delta$AEY$_\text{rel}$} \\
 & \textdegree C & \textdegree C & \textdegree C & \textdegree C & kWh~m$^{-2}$ & kWh~m$^{-2}$ & -- \\
\midrule
& \multicolumn{7}{c}{\textbf{Open-rack mount}}\\
\midrule
Princeton & 39.2 & 36.6 & 4.5 & 2.7 & 420 & 425 & 1.3\%\\
Miami & 44.9 & 42.6 & 4.2 & 2.3 & 447 & 452 & 1.1\%\\
La Paz & 52.3 & 49.4 & 4.6 & 2.9 & 548 & 556 & 1.5\%\\
Golden & 39.4 & 36.7 & 4.7 & 2.7 & 492 & 498 & 1.3\%\\
Ketchikan & 29.5 & 27.3 & 4.5 & 2.2 & 269 & 272 & 1.0\%\\
Berlin & 36.0 & 33.8 & 4.3 & 2.2 & 300 & 303 & 1.0\%\\
\midrule
& \multicolumn{7}{c}{\textbf{Close-roof mount}}\\
\midrule
Princeton & 50.2 & 46.2 & 6.9 & 4.1 & 398 & 406 & 2.1\%\\
Miami & 54.9 & 51.4 & 6.5 & 3.6 & 425 & 433 & 1.9\%\\
La Paz & 64.5 & 60.0 & 7.0 & 4.5 & 513 & 525 & 2.5\%\\
Golden & 51.1 & 46.8 & 7.2 & 4.2 & 464 & 474 & 2.2\%\\
Ketchikan & 38.5 & 35.1 & 6.9 & 3.3 & 258 & 262 & 1.6\%\\
Berlin & 45.5 & 42.1 & 6.6 & 3.4 & 287 & 291 & 1.7\%\\
\bottomrule
\end{tabular}
\end{table*}

\section{Additional results}
\subsection{Results with fixed spectrum}
In the main manuscript we discussed results when spectral irradiance data is available, which are summarized in Table~\ref{tab:results_spectral}. Table~\ref{tab:results_am15} shows the results with the same annual dataset as in Table \ref{tab:results_spectral} but a fixed AM1.5 spectrum and a fixed infrared irradiance fraction of $f_\text{IR}(\mathrm{AM1.5})=12.9\%$. Here, the overall energy yield is estimated to be slightly lower, which is in agreement with 12.9\% being larger than the mean IR fraction for spectral data (see Fig. \ref{fig:histograms}). The relative change in energy yield $\Delta \text{AEY}_\text{rel}$ between PV modules with and without an ideal SBR is a bit larger than when spectral data is available.

\begin{table*}
\caption{Results shown for Princeton for five years (2020--2024) with spectrally variable irradiance data $I_\lambda$ for open-rack and close-roof mounted PV panels, with an SBR cutoff wavelength of $\lambda_\text{SBR}=1150$~nm. The table shows the average temperatures $\overline{T_\text{ref}}$ and  $\overline{T_\text{SBR}}$, the maximal $\Delta T_{\max}=\max[T_\text{ref}(t)-T_\text{SBR}(t)]$, and the average temperature differences $\overline{\Delta T}=\overline{T_\text{ref}}-\overline{T_\text{SBR}}$ of a solar cell in the PV module without (ref) and with ideal sub-bandgap reflector (SBR); the annual energy yield without ($\text{AEY}_\text{ref}$) and with ($\text{AEY}_\text{SBR}$) ideal SBR and the relative change of the energy yield ($\Delta\text{AEY}_\text{rel}=[\text{AEY}_\text{SBR}-\text{AEY}_\text{ref}]/\text{AEY}_\text{ref}$). Results are shown for temperature coefficients of $\beta_V=-0.40\,\%\,\text{K}^{-1}$ and $\beta_J=+0.03\,\%\,\text{K}^{-1}$. Average temperatures were calculated as irradiance-weighted averages, using the plane-of-array irradiance $I_m$ [see eq.\ (\ref{eq:irr_weighted})]. The values of $\overline{\Delta T}$ and $\Delta$AEY$_\text{rel}$ were calculated before rounding and can therefore deviate from values computed from the rounded numbers shown.}
\label{tab:results_annual}
\centering
\begin{tabular}{l|ccccccc}
\toprule
\textbf{Location} & \textbf{$\overline{T_\text{ref}}$} & \textbf{$\overline{T_\text{SBR}}$} & \textbf{$\Delta T_\text{max}$} & \textbf{$\overline{\Delta T}$} & \textbf{AEY$_\text{ref}$} & \textbf{AEY$_\text{SBR}$} & \textbf{$\Delta$AEY$_\text{rel}$} \\
 & \textdegree C & \textdegree C & \textdegree C & \textdegree C & kWh~m$^{-2}$ & kWh~m$^{-2}$ & -- \\
\midrule
& \multicolumn{7}{c}{\textbf{Open-rack mount}}\\
\midrule
2020 & 38.1 & 35.7 & 4.7 & 2.4 & 399 & 403 & 1.1\%\\
2021 & 37.8 & 35.4 & 4.7 & 2.4 & 414 & 419 & 1.1\%\\
2022 & 38.7 & 36.2 & 4.7 & 2.5 & 433 & 438 & 1.2\%\\
2023 & 38.8 & 36.3 & 4.8 & 2.4 & 401 & 406 & 1.1\%\\
2024 & 39.2 & 36.7 & 4.6 & 2.5 & 423 & 428 & 1.2\%\\
\midrule
& \multicolumn{7}{c}{\textbf{Close-roof mount}}\\
\midrule
2020 & 49.0 & 45.3 & 7.2 & 3.7 & 378 & 385 & 1.9\%\\
2021 & 48.7 & 45.0 & 7.2 & 3.7 & 393 & 400 & 1.9\%\\
2022 & 49.9 & 46.0 & 7.2 & 3.9 & 409 & 417 & 2.0\%\\
2023 & 49.6 & 45.9 & 7.2 & 3.7 & 380 & 388 & 1.9\%\\
2024 & 50.2 & 46.5 & 7.0 & 3.8 & 401 & 409 & 1.9\%\\
\bottomrule
\end{tabular}

\end{table*}

\subsection{Annual variability}
In the main manuscript we showed results for single years, because in the NSRDB database spectral data only is available for these datasets. Since annual energy yield depends on year-to-year variations in weather, we tested how sensitive $\Delta\text{AEY}_\text{rel}$ is to these variabilities by evaluating results for Princeton for five consecutive years, 2020--2024, shown in Table~\ref{tab:results_annual}. While the absolute annual energy yield AEY$_\text{ref}$ varies by up to $\sim$8\% (min to max) between years for both mounting configurations, $\Delta\text{AEY}_\text{rel}$ remains stable, ranging from 1.1\% to 1.2\% for open-rack and from 1.9\% to 2.0\% for close-roof mounting. This is because both the reference and SBR modules experience the same year-to-year weather variations, which largely cancel when calculating the relative change $\Delta\text{AEY}_\text{rel}$; only the small residual differences in how each year's spectrum and temperature profile interact with the SBR cutoff remain. We therefore expect our conclusions on the SBR benefit to be robust to interannual weather variability, despite using only a single year of data for each location in the main manuscript.

\section{PV module degradation}
\label{sec:degradation}
We estimate the degradation of PV modules with a simple \emph{Arrhenius} model, where the degradation rate $k$ is given by
\begin{equation*}
    k = \gamma \cdot\exp\left(-\frac{E_a}{k_BT}\right),
\end{equation*}
with the activation energy $E_a$, the temperature $T$ (in kelvin), the Boltzmann constant $k_B$ and a pre-exponential factor $\gamma$. From the energy-yield framework described above we have hourly data for the PV module temperature, which we can use to approximate an annual degradation rate,
\begin{equation}
    \label{eq:k_yr}
    k_\text{yr} = \gamma \cdot\sum_\text{year} \exp\left(-\frac{E_a}{k_BT(t)}\right),
\end{equation}
where the sum is taken over all hours of the year, including night-time, when the Sun is below the horizon. While the different degradation mechanisms in reality may vary over the lifetime of a PV module, we assume that they are constant and that the performance of a PV module follows an exponential decay,
\begin{equation}
    \label{eq:degr_p}
    P(\mathrm{STC},t) = P(\mathrm{STC},t_0) \cdot\exp\left[-k_\text{yr}(t-t_0)\right],
\end{equation}
where $t$ is the time in years. From Eq.\ (\ref{eq:degr_p}) we get the \emph{lifetime} $t_{80\%}$, where the STC power has fallen to 80\% of the initial value, with
\begin{equation}
    t_{80\%} = -\frac{\ln 0.8}{k_\text{yr}}.
\end{equation}

The \emph{cumulative} energy yield $\Sigma$EY is calculated with
\begin{equation}
    \label{eq:ey_cum}
    \Sigma\text{EY}(T) = \sum_{\tau=0}^T \text{AEY}(\tau) = \text{AEY}(0)\sum_{\tau=0}^T(1-k_\text{yr})^\tau,
\end{equation}
where $\tau$ is the time in years, ``0'' corresponds to the first year and $k_\text{yr}$ is the annual degradation rate. The \emph{relative gain in cumulative energy yield} is calculated via

\begin{equation}
    \label{eq:ey_gain}
    \begin{aligned}
    \Gamma(T) &= \frac{\sum_{\tau=0}^T\left[\text{AEY}_\text{SBR}(\tau)-\text{AEY}_\text{ref}(\tau)\right]}{\sum_{\tau=0}^T\text{AEY}_\text{ref}(\tau)}\\
    &=\frac{\Sigma\text{EY}_\text{SBR}(T)-\Sigma\text{EY}_\text{ref}(T)}{\Sigma\text{EY}_\text{ref}(T)}.
    \end{aligned}
\end{equation}

%%%REFERENCES%%%
\makeatletter
\def\bibsection{%
  \par
  \onecolumngrid@push
  \vspace*{10\p@}%
  \begingroup\baselineskip26\p@\bib@device{\textwidth}{245.5\p@}\endgroup
  \nobreak\@nobreaktrue
  \addvspace{19\p@}%
  \par
  \onecolumngrid@pop
}
\makeatother
%\clearpage
\bibliographystyle{apsrev4-2}
\bibliography{references} 

% -------------------------------------------------
\end{document}